\documentclass[a4paper,10pt]{article}
\usepackage{a4wide}
\usepackage[T1]{fontenc}
\usepackage[utf8]{inputenc}
\usepackage{latexsym}
\usepackage{graphicx}

\usepackage{slashed}
\usepackage{amssymb, amsmath, amsfonts}
\usepackage{color}
\usepackage[usenames,dvipsnames,svgnames,table]{xcolor}
\usepackage{a4wide}
\usepackage{mathpazo}
\usepackage{mathrsfs} 
\usepackage{hyperref}
\hypersetup{
    colorlinks,%
    citecolor=blue,%
    filecolor=magenta,%
    linkcolor=red,%
    urlcolor=green
}

\usepackage{txfonts} 

\usepackage{tabularx}
\usepackage[numbers,sort&compress]{natbib}
\usepackage[affil-it]{authblk}
\title{Charged Scalar Pair Production in Strong-Field Photon-Photon Interaction}
\author{M. Onirban Islam, Md. Galib Hassan, and M. Arshad Momen\thanks{Electronic address: \texttt{amomen@univdhaka.edu}}}
\affil{Department of Theoretical Physics, University of Dhaka, Dhaka - 1000, Bangladesh.}

\begin{document}

\maketitle

\begin{abstract}
Following the pioneering work of H. Reiss \cite{Reiss_JMP1962}, we provide a covariant calculation of the charged scalar particle pair 
production. The calculation is facilitated by the use of two-dimensional Bessel functions and light-font coordinates.  
\end{abstract}

\section{Introduction}
\label{intro}
After the recent discovery of Higgs like particles \cite{ATLAS_Collaboration_PLB2012, CMS_Collaboration_PLB2012} Large Hadron Collider (LHC)  one would 
be necessarily looking forward to the design and construction of the next generation of colliders 
\cite{Hesselbach_JPConfSer2009, Jaeckel_JPConfSer2009, Gies_EPJD2009}. It is apparent 
for practical reasons ( both in terms of resources and technology) to concentrate for processes where 
initial number particles is high rather than just high in individual energy. One such experimentally verified 
process~\cite{Bamber_PRD1997, Burke_PRL1997, Hu_PRL2010} is the production of massive charged particles by 
the collision of photons, known as the Breit-Wheeler process \cite{BreitWheeler_PR1934}.
This sort of process involves uncharged particles like photons thus allowing one to focus them in a beam accurately as well as detecting the charged particles produced in the 
final state due to the presence of a clean background. As shown by one of the founding fathers of QED \cite{Schwinger_PR1951} for sufficiently strong electric fields 
one can have nonperturbative effects like pair production of charged fermions. However, even before reaching the Schwinger limit, other 
nonperturbative effects can occur, like the multi-photon Breit-Wheeler process \cite{Reiss_JMP1962, Reiss_PRL1971, 
NikishovRitus_JETP1964a, NikishovRitus_JETP1964b, NikishovRitus_JETP1965, BrownKibble_PR1964}. Since then,  
intensive theoretical research has been undertaken to study the SFQED processes  
\cite{MarklundShukla_RMP2006,Ehlotzky_RPP2009,Ruffini_PR2010,Piazza_RMP2012}.

Though the interactions between fermions and intense electromagnetic fields have been investigated intensively, due to the absence of 
experimental detection of fundamental (charged) scalars, similar processes in scalar electrodynamics have not been that numerous. 
To the best of our knowledge, only Cheng and Wu 
\cite{ChengWu_PRL1969_22, ChengWu_PRL1969_23, ChengWu_PR1969_i, ChengWu_PR1969_ii, ChengWu_PR1969_iii, 
ChengWu_PR1969_iv, ChengWu_PRD1970_p3414, ChengWu_PRD1970_p467, ChengWu_PRD1970_2, Ruffini_PR2010} 
have investigated the pion
production using fourth order in \textit{perturbation} theory. But as mentioned above, the current SF regime (as exemplified in Fig. 5 \cite{Hu_PRL2010}) 
is no more in the arena of perturbation 
theory. In this present work we look at the pair production of charged scalar pair production by photon-photon interaction also, but, 
using the nonperturbative approach of Reiss \cite{Reiss_JMP1962}, who had investigated the mutual absorption 
of two plane electromagnetic waves in a fashion that one of the fields is treated accurately, named as the background field  
and the other one is considered as a perturbation. This approach, closer to the spirit of relativistic quantum mechanics (RQM), has been 
shown to be equivalent to laser-induced phenomena \cite{FriedEberly_PR1964, Reiss_LP2005, Reiss_EPJD2009}.

\begin{figure}
\centering
  \includegraphics[height=5cm]{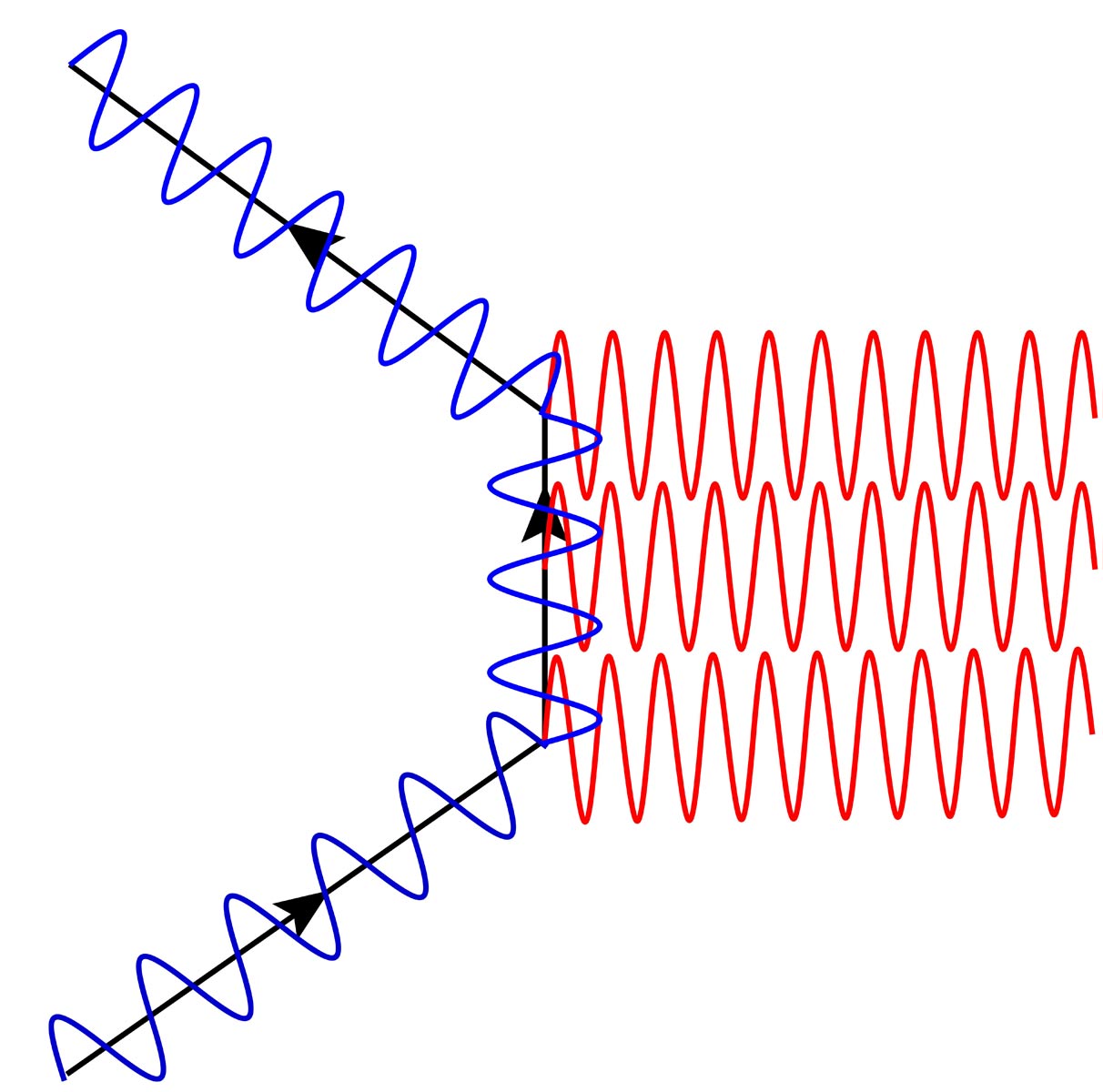}
\caption{Strong-field Breit-Wheeler process.}
\label{fig: BW}       
\end{figure}

Our note starts with a summary of Gordon-Volkov solution \cite{Gordon_ZP1926, Volkov_ZPhys1935} in sec. \ref{sec: 1}, 
then the $S$-matrix element is calculated in sec.  \ref{sec: 2}, followed by the production rate in sec. \ref{sec: 3}.

\section{Klein-Gordon Field in Strong External EM Field} \label{sec: 1}
The Klein-Gordon equation with minimul classical electromagnetic field coupling is
\begin{equation} \label{eq: KGEM}
	(\mathrm{D}^2 + m^2) \phi = 0, \; \mathrm{D}_{\mu} \equiv \partial_{\mu} - i e A_{\mu} (x),
\end{equation}
where $m$ is the mass of the particle, $A (x)$ is electromagnetic potential, and $e$ is the charge of the electrodynamic field. 
The exact solution of Eq. (\ref{eq: KGEM}) normalized in quantized volume $V$ for a plane electromagnetic wave $A(\xi), \xi \equiv k \cdot x$, where $k$ being the wave 
vector of the electromagnetic field, was obtained by Gordon \cite{Gordon_ZP1926, Ehlotzky_RPP2009, Boca_JPA2011}
\begin{equation} \label{eq: GordonSoln}
	\phi_\pm (p; x) = \frac{1}{ \sqrt{2EV}} \exp[\mathrm{i}  I],
\end{equation}
where $E$ is the time component of the momentum eigenvalue and $I$ stands for the action of a 
classical particle in a plane wave (damped at infinity) \cite{Berestetskii_Elsevier1982} :
\begin{equation} \label{eq: actionVolkov}
I = - p \cdot x - \int_0^{k \cdot x} \mathrm{d} \xi 
\biggr[ \frac{e}{k \cdot p} A (\xi) \cdot p - \frac{e^2}{2 k \cdot p} A^2 (\xi) \biggr]
\end{equation}
The action of a classical particle interacting with a plane electromagnetic wave can be written compactly as
\begin{equation}
	I = - q \cdot x,
\end{equation}
where $q$ is the quasimomentum of the particle in the field \cite{Volkov_ZPhys1935, SenGupta1952, Ritus_1985}
\begin{equation}
	q_\mu \equiv p_\mu + N k_\mu - n k_\mu.
\end{equation}
Looking at the defining expression of $Q$, we can easily interpret the following quantities as some numbers $N, n \in \mathbb{Z}^{+}$:
\begin{equation} \label{eq: def_N_n}
	N \equiv \frac{e}{k \cdot p} \langle A \rangle \cdot p, \quad n \equiv \frac{e^2 \langle A^2 \rangle }{2 k \cdot p}.
\end{equation}
It satisfies \cite{SenGupta1952, Reiss_JMP1962, NikishovRitus_JETP1964a, NikishovRitus_JETP1964b, 
NikishovRitus_JETP1965, BrownKibble_PR1964, Goldman_a, Berestetskii_Elsevier1982, Ritus_1985, Reiss_EPJD2009, Piazza_RMP2012} 
\begin{equation} \label{eq: onShell_dressedElectron}
	q^2 = M^2,
\end{equation}
where $M$ is the renormalized mass of the electron in the field, given by \cite{Reiss_JMP1962, Ritus_1985,  Reiss_PQE1992}:
\begin{equation} \label{eq: massShift}
	M = m \sqrt{1 + \varXi^2},
\end{equation}
where the Lorentz \cite{Reiss_PQE1992} and gauge \cite{HeinzlIlderton_OptCommun2009} invariant  definition of the classical 
nonlinearity parameter is
\begin{equation}
	\varXi \equiv \frac{e^2}{m^2} \frac{\langle p_\mu T^{\mu\nu} p_\nu \rangle}{(k \cdot p)^2} = \frac{e \sqrt{ \langle A^2 \rangle } }{m},
\end{equation}
where $T^{\mu \nu}$ is the Maxwell energy-momentum tensor.
\newline

Due to the interaction with the background field,  propagating assumed along the $x^{3}$ direction, we have
\begin{equation}
	[ \mathrm{p}_{\varpi}, \mathrm{H} ] \neq 0, \quad \varpi = 0, 3,
\end{equation}
but
\begin{equation}
	[ \mathrm{p}_{0} - \mathrm{p}_{3}, \mathrm{H} ] = 0, \;\textrm{and}\; (\mathrm{p}_{0} - \mathrm{p}_{3}) | \phi \rangle = \rho | \phi \rangle.
\end{equation}
In the limit of $A \to 0$, the last equation gives
\begin{equation*}
	\rho = \pm |E| - p_3.
\end{equation*}
Here $E$ and $p_3$ have no physical significance unless we are in the limit $A \to 0$, where they will have the usual physical 
interpretation as energy and third component of momentum \cite{Berestetskii_Elsevier1982}.

\section{Calculation of $S$-matrix} \label{sec: 2}
Let us now concentrate on the scattering process schematically shown in Fig. \ref{fig: BW}. The blue and red lines represent the 
background and the perturbing fields respectively. For the perturbing field $A'(\xi') = \varepsilon'_{\mu} f' (\xi') $ in Lorentz gauge, 
the transition amplitude is \cite{Reiss_PQE1992}
\begin{eqnarray}
	T_{\rm fi} & = & - \mathrm{i} \Big( \phi^*(p_{\rm f}), \, [2 \mathrm{i} e A' \cdot \partial - e^2 A'^2] \phi (p_{\rm i}) \Big) \nonumber\\
& = & \frac{- \mathrm{i} e}{ 2 V \sqrt{E_{\rm f} E_{\rm i}}} (\mathscr{N}_1 + \mathscr{N}_2) \label{eq: transitionExprGenKG},
\end{eqnarray}
where the numbers $\mathscr{N}$'s are defined as
\begin{subequations}
   \begin{eqnarray}
	\mathscr{N}_1(t) & \equiv & -  \varepsilon' \cdot p_{\rm i} \int \mathrm{d}^4 x 	\exp[\mathrm{i} \mathfrak{B}] f' (\xi')   \label{eq: N1_KG_def}\\
	\mathscr{N}_2 (t) & \equiv & e |\varepsilon'|^2 \int \mathrm{d}^4 x \exp[\mathrm{i} \mathfrak{B}] f'^2 (\xi')  \label{eq: N2_KG_def} 
   \end{eqnarray}
\end{subequations}
and $\mathfrak{B}$ is defined by 
\begin{equation} \label{eq: Bristy}
      \mathfrak{B} \equiv- (I_{\rm f} - I_{\rm i}) = ( q_{\textrm{f}} - q_{\textrm{i}} ) \cdot x.
\end{equation}
Here we have used the transversality condition $k \cdot \varepsilon' = 0$ \cite{Berestetskii_Elsevier1982} and the gauge transformation 
for the polarization vector of the perturbing field \cite{BrownKibble_PR1964, Ritus_1985}
\begin{equation}
	\varepsilon'_{\mu} \longrightarrow \varepsilon'_{\mu} - \frac{k \cdot \varepsilon'}{k \cdot k'} k'_{\mu}
\end{equation}
due to the gauge freedom \cite{Berestetskii_Elsevier1982}. Up to now, we have derived the equations for a general 
$A_{\mu} = \varepsilon_{\mu} f, A'_{\mu} = \varepsilon'_{\mu} f'$ profile and hence, 
is applicable for any scalar elementary relativistic particles. Now we choose a specific profile representing the experimental set-up of the 
background field propagating along the $x^{3}$ direction and the perturbing one being the opposite:
$k_{\mu} = (\omega, 0, 0, \omega), k'_{\mu} = (\omega, 0, 0, - \omega), f = a \cos\xi, \, f' = a' \exp[ - \mathrm{i}\xi']$. In this case, we obtain
\begin{eqnarray} \label{eq: BMinus_zeta'}
	\mathfrak{B} - \xi' & = & (\rho_{\rm f} - \rho_{\rm i} - 2 \omega') t - (\boldsymbol{p}^{\perp}_{\textrm{f}} - \boldsymbol{p}^{\perp}_{\textrm{i}}) \cdot \boldsymbol{x}^{\perp} 
	+ \mathscr{A} \sin\xi + \mathscr{B} \sin2\xi + \mathscr{C} \xi, 
\end{eqnarray}
where
\begin{subequations}
	\begin{eqnarray}
		\mathscr{A} & = &  \frac{ae}{\omega} \left( \frac{\varepsilon \cdot p_{\rm f}}{\rho_{\rm f}} - \frac{\varepsilon \cdot p_{\rm i}}{\rho_{\rm i}} \right),\\
\mathscr{B} & = & \frac{a^2 e^2}{4 \omega} \left( \frac{1}{\rho_{\rm f}} - \frac{1}{\rho_{\rm i}} \right),\\
\mathscr{C}_{1} & = & \frac{a^2 e^2}{4 \omega}   \left( \frac{1}{\rho_{\rm f}} - \frac{1}{ \rho_{\textrm{i}} } \right) + \frac{(p_{\textrm{f}_{3}} - p_{\textrm{i}_3})}{\omega} \xi  + \frac{\omega'}{\omega} .
	\end{eqnarray}
\end{subequations}
Substituting the expression of $\mathfrak{B} - \xi'$ to get $\mathscr{N}$, it is required to set integration limit. We introduce two 
artificial length parameters, $L$ and $L_{\perp} = ( - \infty, \infty)$ and define volume $V \equiv L L_{\perp}^2$, and with the interaction 
time $t$ we have for $\mathscr{N}_{1}$:
\begin{eqnarray}
	\mathscr{N}_1 & = & - a' \varepsilon' \cdot p_{\textrm{i}} 
	\int_0^t \exp[ \mathrm{i} (\rho_{\rm f} - \rho_{\rm i} - 2 \omega') \grave{t}] \mathrm{d} \grave{t}
	 \int_{- \infty}^{\infty} \exp[ - \mathrm{i} ( \boldsymbol{p}_{\mathrm{f}}^{\bot} - \boldsymbol{p}_{\mathrm{i}}^{\bot}) \cdot \boldsymbol{x}^{\bot}] \mathrm{d} \boldsymbol{x}^{\bot}\nonumber\\
	 && \sum \int_{-L/2}^{L/2} \exp\left[ \mathrm{i} (\mathscr{A} \sin\xi + \mathscr{B} \sin2\xi + \mathscr{C}_{1} \xi) \right] \mathrm{d} x^3 \nonumber\\
	& = & - 2 \pi \delta^{2} \left(\boldsymbol{p}^{\perp}_{\textrm{f}} - \boldsymbol{p}^{\perp}_{\textrm{i}} \right) a' \varepsilon' \cdot p_{\rm i} 
	 \frac{\exp[ \mathrm{i} (\rho_{\textrm{f}} - \rho_{\textrm{i}} - 2 \omega') t] - 1}{\mathrm{i} (\rho_{\textrm{f}} - \rho_{\textrm{i}} - 2 \omega')} 
	 \sum \int_{-L/2}^{L/2} \exp\left[ \mathrm{i} (\mathscr{A} \sin\xi + \mathscr{B} \sin2\xi + \mathscr{C}_{1} \xi) \right] \mathrm{d} x^3. \nonumber
\end{eqnarray}
The integral over $x^3$ becomes independent of $t$ if $ \mathscr{C}_{1}, \mathscr{C}_{2} = n \in \mathbb{Z} $ and the implementation 
of this is $\omega' L = 2 \pi \jmath, \; \jmath \in \mathbb{Z}$, compatible with the box quantization, so that, $\omega'$ 
must satisfy the same periodicity condition as $\omega$. For convenience, we transform $\xi$ to another variable $\theta$ \cite{Reiss_JMP1962}, 
and we have for $\mathscr{N}_{1}$ and $\mathscr{N}_{2}$
\begin{eqnarray}
	\mathscr{N}_1 & = &  2 \pi a' \varepsilon' \cdot p_{\textrm{i}} \delta^{2} (\boldsymbol{p}^{\perp}_{\textrm{f}} - \boldsymbol{p}^{\perp}_{\textrm{i}} ) 
	\frac{\exp[ \mathrm{i} (\rho_{\textrm{f}} - \rho_{\textrm{i}} - 2 \omega') t] - 1}{\mathrm{i} (\rho_{\textrm{f}} - \rho_{\textrm{i}} - 2 \omega')} \sum_{n} \mathscr{J}_{n} (\mathscr{A}, \mathscr{B}), \\
		  \mathscr{N}_2 & = &  2 \pi e a'^{2}|\varepsilon'|^2 \delta^{2} (\boldsymbol{p}^{\perp}_{\textrm{f}} - \boldsymbol{p}^{\perp}_{\textrm{i}} ) 
	\frac{\exp[ \mathrm{i} (\rho_{\textrm{f}} - \rho_{\textrm{i}} - 4 \omega') t] - 1}{\mathrm{i} (\rho_{\textrm{f}} - \rho_{\textrm{i}} - 4 \omega')} \sum_{n} \mathscr{J}_{n} (\mathscr{A}, \mathscr{B}),
\end{eqnarray}
where we have used the integral representation of two-dimensional Bessel function $\mathscr{J}_{n}(x, y)$ \cite{Korsch_JPA2006}.

\section{Production rate} \label{sec: 3}
In the present work, the states with quantum numbers $\boldsymbol{p}, p_{0} = - \sqrt{m^2 + \boldsymbol{p}^2}$ propagates backward 
in the time and is interpreted as a positron with quantum numbers $ - \boldsymbol{p}, - p_{0} = \sqrt{m^2 + \boldsymbol{p}^2} > 0$, 
propagates forward in time. Therefore, the creation of a positron with ``momentum'' 
$p_{\mathrm{i}_\mu} = \left( \sqrt{m^2 + \boldsymbol{p}_{\mathrm{i}}^2}, \boldsymbol{p}_{\mathrm{i}} \right)$ and of an electron with 
``momentum'' $p_{\mathrm{f}_\mu} = \left( \sqrt{m^2 + \boldsymbol{p}_{\mathrm{f}}^2}, \boldsymbol{p}_{\mathrm{f}} \right)$
 described as a transition of an electron from a state with quantum number $- p_{\mathrm{i}} $ into a state with quantum number 
$p_{\mathrm{f}}$ \cite{Ritus_1985},
as depicted by the Feynman diagram \ref{fig: BW}. Thus for particles with zero spin, the total probability per unit time per unit 
volume for creating a pair is 
\begin{equation} \label{eq: productionRateWithOutSpin}
	\mathrm{d} P = \lim_{t \to \infty} \frac{1}{t}
V \widetilde{\mathrm{d} p_{\mathrm{f}}} V \widetilde{\mathrm{d} p_{\mathrm{i}}} |T_{\mathrm{fi}}|^2,
\end{equation}
where $\widetilde{\mathrm{d} p}$ is the Lorentz invariant measure of the phase space in per unit volume $V$.
To get the production rate, we need
\begin{equation*}
	|T_{\textrm{fi}}|^2 = \frac{e^2}{ 4 V^{2} E_{\textrm{f}} E_{\textrm{i}} } \left( |\mathscr{N}_1|^2 + |\mathscr{N}_2|^2 + 2 \Re \mathscr{N}_1^{*} \mathscr{N}_2 \right).
\end{equation*}
To perform the limit $t \to \infty$, we use the integral representation of Dirac delta function \cite{Berestetskii_Elsevier1982}
\begin{eqnarray}
	\lim_{t \to \infty} \frac{1}{t} \frac{2 [1 - \cos(\rho_f - \rho_i - 2 \omega')t]}{(\rho_f - \rho_i - 2 \omega')^2} 
	& = & 2 \pi \delta (\rho_f - \rho_i - 2 \omega'),
\end{eqnarray}
After taking the $t \to \infty$ limit, we obtain
\begin{subequations}
	\begin{eqnarray}
		\lim_{t \to \infty}|\mathscr{N}_1|^2 & = & 8 \pi^3 (a' \varepsilon' \cdot p_{\rm i})^{2} \delta (\rho_{\textrm{f}} - \rho_{\textrm{i}} - 2 \omega') \delta^2 (\boldsymbol{p}^\bot_{\textrm{f}} - \boldsymbol{p}^\bot_{\textrm{i}}) \delta^2 (\boldsymbol{p}'^\bot_{\textrm{f}} - \boldsymbol{p}^\bot_{\textrm{i}}) \mathscr{J}^2, \label{eq: N1_KG_final}\\
	        \lim_{t \to \infty}|\mathscr{N}_2|^2 & = & 8 \pi^3 e^2 a'^{4}|\varepsilon'|^4  \delta (\rho_{\textrm{f}} - \rho_{\textrm{i}} - 4 \omega') \delta^2 (\boldsymbol{p}^\bot_{\textrm{f}} - \boldsymbol{p}^\bot_{\textrm{i}}) \delta^2 (\boldsymbol{p}'^\bot_{\textrm{f}} - \boldsymbol{p}^\bot_{\textrm{i}}) \mathscr{J}^2 \label{eq: N2_KG_final},
	\end{eqnarray}
\end{subequations}
where $\mathscr{J}^{2} = |\sum_{n} \mathscr{J}_{n}|^2$ and the cross term $\mathscr{N}_1^{*} \mathscr{N}_2$ becomes zero. 
The products of the $\delta$ distribution of $\boldsymbol{p}^{\perp}$ is zero, unless $\boldsymbol{p}'^\bot_{\textrm{f}} = \boldsymbol{p}^\bot_{\textrm{f}}$ 
and then, we have
\begin{equation}
	\delta^2 (\boldsymbol{p}^\bot_{\textrm{f}} - \boldsymbol{p}^\bot_{\textrm{i}}) \delta^2 (\boldsymbol{p}'^\bot_{\textrm{f}} - \boldsymbol{p}^\bot_{\textrm{i}}) 
	=
	\delta^2 (\boldsymbol{p}^\bot_{\textrm{f}} - \boldsymbol{p}^\bot_{\textrm{i}}) \frac{\mathcal{A}}{(2 \pi)^2}, \quad \mathcal{A} = \int_{- \infty}^{\infty} \mathrm{d} \boldsymbol{x}^{\perp} = L_{\perp}^{2},
\end{equation}
i.e., proportional to an infinite area in the perpendicular plane of electromagnetic beams. We calculate the production rate in per unit area: 
$\mathscr{P} = P / \mathcal{A}$.
Putting Eqs. (\ref{eq: N1_KG_final}) and (\ref{eq: N2_KG_final}) into the Eq. (\ref{eq: productionRateWithOutSpin}) 
and after taking the $t$ limit, we have in the CM frame for the production rate
\begin{eqnarray} \label{eq: integral_productionRate}
	\mathscr{P} = \int \widetilde{\mathrm{d} p_{\rm{f}}} \widetilde{\mathrm{d} p_{\rm{i}}}  \delta^{2} (\boldsymbol{p}^\bot_{\textrm{f}} - \boldsymbol{p}^\bot_{\textrm{i}}) 8 \pi^{3} \mathscr{J}^2 (\mathscr{A}, \mathscr{B})
\biggr[ \delta (\rho_{\rm f} - \rho_{\rm i} - 2 \omega') (a' \varepsilon' \cdot p_{\rm i})^2
+ e^2 a'^{4} |\varepsilon'|^4 \delta (\rho_{\rm f} - \rho_{\rm i} - 4 \omega') \biggr].
\end{eqnarray}

Physically the Dirac delta function represents the energy conservation of the multi-photon process
\begin{equation}
	E_{\textrm{f}} - E_{\textrm{i}} = ( n_{\textrm{f}} - n_{\textrm{i}} ) \omega + \omega' = n \omega + \omega',
\end{equation}
where $n_{\textrm{f}}, n_{\textrm{i}}$ are defined by Eq. (\ref{eq: def_N_n}) and these are conserved quantities for a particular laser 
parameter $\varXi$ giving the number of photon exchange to ensure particle production
\begin{equation}
	n \omega + \omega' \geq 2 M.
\end{equation}
This condition together with the momentum conservation in center of mass reference frame $n \omega = \omega' \geq M$ gives
\begin{equation} \label{eq: n_minCondition}
	n \omega \omega' \geq M^{2},
\end{equation}
from which we can deduce the minimum number of required photons
\begin{equation} \label{eq: n_min}
	n_{\rm min} = \left\lfloor \frac{M^2}{\omega\omega'} \right\rfloor,
\end{equation}
where $\lfloor \hdots \rfloor$ symbolizes the largest integer. Since, the inner product of propagation vectors is a Lorentz scalar of value
\begin{equation*}
	k \cdot k' = \omega\omega' - \boldsymbol{k} \cdot \boldsymbol{k}' = 2 \omega\omega',
\end{equation*}
inequality (\ref{eq: n_minCondition}) and Eq. (\ref{eq: n_min}) holds in any relativistic frame.
\newline

The integral (\ref{eq: integral_productionRate}) cannot be evaluated in a straightforward way due to the complicated dependency of different components of momentum 
vector inside the delta function. This sort of complicated dependency is a consequence of a QED field interacting with another background 
field and arises in different fundamental processes 
\cite{Reiss_JMP1962, BrownKibble_PR1964, Berestetskii_Elsevier1982, Ritus_1985, Ehlotzky_RPP2009, Piazza_RMP2012}.
The evaluation of this sort integrals has been performed using the method of steepest descent in early days of SFQED, e.g. 
\cite{Reiss_JMP1962, Ritus_1985} or by some algebraic manipulation inside the delta distribution \cite{BrownKibble_PR1964}, and 
recently subjected to multi-dimensional Monte Carlo integration \cite{Ehlotzky_RPP2009, Hu_PRL2010}.
To perform the above integral, it is convenient to use light-cone form of relativistic dynamics \cite{Dirac_RMP1949}. 
We place a tilde over a quantity to represent its light-cone 
coordinates representation, so the spacetime and momentum variables in this coordinate system are $\tilde{x}^\mu = (x^+, x^-, 
\boldsymbol{x}^\bot), \boldsymbol{x}^\bot = (x^2, x^3), \mu = +, -, 2, 3$ and $\tilde{p}_\mu = (p_+, p_-, \boldsymbol{p}_\bot), 
\boldsymbol{p}_\bot = (p_1, p_2)$, where the time $x^+$ and longitudinal space $x^-$ variables are defined as
\begin{equation*}
	x^{\pm} \equiv x^0 \pm x^3, p^{\pm} \equiv p^0 \pm p^3
\end{equation*}
and the phase space measure is
\begin{equation*}
	\widetilde{\mathrm{d} p} = \int_{- \infty}^{\infty} \frac{\mathrm{d}^2 \boldsymbol{p}^{\perp}}{2 \pi} \int_{0}^{\infty} \frac{\mathrm{d} p^{-} }{ \sqrt{4 \pi p^{-} } }.
\end{equation*}
Rewriting the production rate in this coordinates with the initial conditions $\varepsilon^{+} = 0 = p_{\textrm{i}}^{+}$, we have
\begin{eqnarray}
	\mathscr{P} 
& = & \int  \mathrm{d} \boldsymbol{p}^{\bot} \frac{\mathrm{d} p_{\mathrm{i}}^{-} }{  2 \sqrt{ p_{\rm i}^{-} }  } 
\left[ a'^{2} |\boldsymbol{p}^{\bot}|^2 \cos^2\varTheta_{\varepsilon}   \frac{\mathscr{J}^2 (\mathscr{A}_{2}, \mathscr{B}_{2}) }{ \sqrt{p_{\rm i}^{-} + 2 \omega'} } 
 + e^2 a'^{4} \frac{\mathscr{J}^2 (\mathscr{A}_{4}, \mathscr{B}_{4}) }{ \sqrt{p_{\rm i}^{-} + 4 \omega'} }  \right],
\end{eqnarray}
where
\begin{subequations}
	\begin{eqnarray}
		\mathscr{A}_{2} & = & \frac{ae}{\omega} \left( \frac{1}{p_{\textrm{i}}^{-}}
- \frac{1}{p_{\textrm{i}}^{\bot} + 2 \omega'} \right) \boldsymbol{\varepsilon}^\bot \cdot \boldsymbol{p}^{\bot},\\
\mathscr{B}_{2} & = & \frac{a^2 e^2}{4 \omega} \left( \frac{1}{p_{\textrm{i}}^{-} + 2 \omega'} - \frac{1}{p_{\textrm{i}}^{-}} \right),\\
\mathscr{A}_{4} & = & \frac{ae}{\omega} \left( \frac{1}{p_{\textrm{i}}^{-}}
- \frac{ 1}{p_{\textrm{i}}^{-} + 4 \omega'} \right) \boldsymbol{\varepsilon}^\bot \cdot \boldsymbol{p}^{\bot},\\
\mathscr{B}_{4} & = & \frac{a^2 e^2}{4 \omega} \left( \frac{1}{p_{\textrm{i}}^{-} + 4 \omega'} - \frac{1}{p_{\textrm{i}}^{-}} \right),
	\end{eqnarray}
\end{subequations}
and $\varTheta_{\varepsilon'}$ is the angle between $\boldsymbol{\varepsilon}'^{\perp}$ and $\boldsymbol{p}^{\perp}$
and we have used $|\boldsymbol{\varepsilon}^\bot| = 1$ as a consequence of $\varepsilon^2 = - 1$. 
To carry out the $\mathrm{d} \boldsymbol{p}_{\mathrm{f}\bot}$ integration, we use the polar coordinates:
\begin{equation*}
	\mathrm{d} \boldsymbol{p}_{\mathrm{f}}^{\bot} = | \boldsymbol{p}_{\mathrm{f}}^{\bot} | \;
\mathrm{d} | \boldsymbol{p}_{\mathrm{f}}^{\bot} | \mathrm{d} \theta_{12},
\end{equation*}
where $\theta_{12}$ is the angles between $\boldsymbol{p}_{\textrm{f}_{1}}$ and $ \boldsymbol{p}_{\textrm{f}_{2}} $.
The integration over $\mathrm{d} \theta_{12}$ gives $2 \pi$. The remaining radial part cannot be evaluated in a straightforward way. 
We observe that only the first arguments of the two-dimensional Bessel functions depend on the radial part:
\begin{equation}
	\mathscr{A}_{4} = c_4 |\boldsymbol{p}^{\bot}|, \quad c_4 =  \frac{ae}{\omega} \left( \frac{1}{p_{\textrm{i}}^{-}}
- \frac{1}{p_{\textrm{i}}^{-} + 4 \omega'} \right)  \cos \varTheta_{\varepsilon},
\end{equation}
and $\varTheta_{\varepsilon}$ is the angle between $\boldsymbol{\varepsilon}^{\perp}$ and $\boldsymbol{p}^{\perp}$. 
So we decompose the two-dimensional Bessel functions in terms of ordinary Bessel functions $J_{n} (x)$ \cite{Korsch_JPA2006}:
\begin{eqnarray} \label{eq: decompose_2DBF}
	 \int_{M^2}^{n \omega \omega'} |\boldsymbol{p}_{\bot}|\; \mathrm{d} |\boldsymbol{p}^{\bot}| \mathscr{\mathscr{J}}^2_{n} (\check{\mathscr{A}}, \check{\mathscr{B}}) 
= \sum_{k, k' = - \infty}^{\infty} \int_{M^2}^{n \omega \omega'} |\boldsymbol{p}^{\bot}| 
 J_{n - 2 k} (c_4|\boldsymbol{p}^{\bot}|) J_{n - 2 k'} (c_4|\boldsymbol{p}^{\bot}|) \, \mathrm{d} |\boldsymbol{p}^{\bot}|\;  
 J_{k} \left( \frac{ea}{4} c_{4} \right) J_{k'} \left( \frac{ea}{4} c_{4} \right).
\end{eqnarray}
For strong-field physics, $\varXi > 1$, the dressed mass can be well approximated by $M^2 \approx m^2 \varXi^2$, and thus, the minimum 
number of photon exchange (\ref{eq: n_min}) is given by 
$n_{\textrm{min}} \approx m^2 \varXi^2 / \omega \omega'$. This makes the upper limit an integral multiple of the lower limit,
$\eta M^2, \eta \equiv n / n_{\textrm{min}} > 1$. Thus for strong-field regime, we have
\begin{equation}
	 \int_{M^2}^{n \omega \omega'} |\boldsymbol{p}^{\bot}|\; \mathrm{d} |\boldsymbol{p}^{\bot}| \mathscr{\mathscr{J}}^2_{n} (\mathscr{A}_{2}, \mathscr{B}_{2}) 
	 \approx \int_{M^2}^{\eta M^{2}} |\boldsymbol{p}^{\bot}|\; \mathrm{d} |\boldsymbol{p}^{\bot}| \mathscr{\mathscr{J}}^2_{n} (\mathscr{A}_{2}, \mathscr{B}_{2}) \nonumber
\end{equation}
and the integration over only the $\boldsymbol{p}^{\perp}$ gives
\begin{eqnarray} \label{eq: diff_P2}
	 \int_{M^2}^{n \omega \omega'} |\boldsymbol{p}^{\bot}| 
 J_{n} (c_{2}|\boldsymbol{p}^{\bot}|) J_{m} (c_{2}|\boldsymbol{p}^{\bot}|) \, \mathrm{d} |\boldsymbol{p}^{\bot}| = 
	 M^{2} \Gamma (N_{1}) \Gamma (N_{2} / 2) (c_{2} M^{2})^{N_{0}} {}_{3}\bar{F}_{4} (C; M^{4}) - \eta^{N_{2}} {}_{3}\bar{F}_{4} (C; \eta^{2} M^{2}),
\end{eqnarray}
where ${}_{3}\bar{F}_{4}$ is the hypergeometric regularized function, the indices $n$ and $m$ represent $n - 2 k$ and $n - 2 k'$ 
respectively and we have used the short hands: 
$N_{i} = n + m +i, C = N_{1} / 2, N_{2} / 2, N_{2} / 2; m + 1, N_{4} / 2, n + 1, N_{1}$ . The integration over 
$\mathrm{d}|\boldsymbol{p}^{\bot}|$ for the term containing $\mathscr{J} (\mathscr{A}_{4}, \mathscr{B}_{4} )$ gives
\begin{eqnarray} \label{eq: diff_P4}
	 \int_{M^2}^{n \omega \omega'} |\boldsymbol{p}^{\bot}|^{3} 
 J_{n} (c_{4}|\boldsymbol{p}^{\bot}|) J_{m} (c_{4}|\boldsymbol{p}^{\bot}|) \, \mathrm{d} |\boldsymbol{p}^{\bot}| = 
	 M^{4} \Gamma (N_{1}) \Gamma (N_{4} / 2) (c_{2} M^{2})^{N_{0}} {}_{3}\bar{F}_{4} (C'; M^{4}) - \eta^{N_{4}} {}_{3}\bar{F}_{4} (C'; \eta^{2} M^{2}),
\end{eqnarray}
where $ C = N_{1} / 2, N_{2} / 2, N_{4} / 2; m + 1, N_{6} / 2, n + 1, N_{1} $. 
The expressions~(\ref{eq: decompose_2DBF}, \ref{eq: diff_P2}, \ref{eq: diff_P4}) gives the differential production rate 
$\mathrm{d} \mathscr{P} / \mathrm{d} p^{-}$ of charged scalar pair production.



\section{Conclusion} \label{eq: conclusion}
We have derived the charged scalar particle production rate in strong-field Breit-Wheeler process. The evaluation of this integral is numerically tenuous and time consuming. Though we have not evaluated these numbers exactly, the fact that this rate can be evaluated in a closed form is encouraging enough to motivate us to persue this calculation further. We hope to report on this in the near future.


\begin{thebibliography}{99}

\bibitem{Reiss_JMP1962}
H.R. Reiss, J. Math. Phys. 3, 59 (1962)

\bibitem{ATLAS_Collaboration_PLB2012}
ATLAS Collaboration, Phys. Lett. B 716, 1 (2012)

\bibitem{CMS_Collaboration_PLB2012}
CMS Collaboration, Phys. Lett. B 716, 30 (2012)


\bibitem{Hesselbach_JPConfSer2009}
S. Hesselbach, J. Phys. Conf. Ser. 198, 012001 (2009)

\bibitem{Jaeckel_JPConfSer2009} 
J. Jaeckel, J. Phys. Conf. Ser. 198, 102008 (2009)

\bibitem{Gies_EPJD2009}
H. Gies, Eur. Phys. J. D 55(2), 311 (2009)

\bibitem{Bamber_PRD1997}
C. Bamber, S.J. Boege, T. Koffas, T. Kotseroglou, A.C. Melissinos, D.D. Meyerhofer, D.A. Reis,
W. Ragg, C. Bula, K.T. McDonald, E.J. Prebys, D.L. Burke, R.C. Field, G. Horton-Smith, J.E.
Spencer, D. Walz, S.C. Berridge, W.M. Bugg, K. Shmakov, A.W. Weidemann, Phys. Rev. D 60,
092004 (1997)

\bibitem{Burke_PRL1997}
D. L. Burke, R. C. Field, G. Horton-Smith, J. E. Spencer, D. Walz, S. C. Berridge, W. M. Bugg, K.
Shmakov, A. W. Weidemann, C. Bula, K. T. McDonald, E. J. Prebys, C. Bamber, S. J. Boege, T.
Koffas, T. Kotseroglou, A. C. Melissinos, D. D. Meyerhofer, D. A. Reis, and W. Raggk, Phys. Rev.
Lett. 79(9), 1626 (1997)

\bibitem{Hu_PRL2010}
H. Hu, C. Müller, C.H. Keitel, Phys. Rev. Lett. 105(8), 080401 (2010)


\bibitem{BreitWheeler_PR1934}
G. Breit, J.A. Wheeler, Phys. Rev. 46, 1087 (1934)

\bibitem{Schwinger_PR1951}
 J. Schwinger, Phys. Rev. 82, 664 (1951)

\bibitem{Reiss_PRL1971}
H.R. Reiss, Phys. Rev. Lett. 26(7), 1072 (1971)

\bibitem{NikishovRitus_JETP1964a}
A.I. Nikishov, V.I. Ritus, Sov. Phys. JETP 19, 529 (1964)

\bibitem{NikishovRitus_JETP1964b}
A.I. Nikishov, V.I. Ritus, Sov. Phys. JETP 19, 1191 (1964)

\bibitem{NikishovRitus_JETP1965}
A.I. Nikishov, V.I. Ritus, Sov. Phys. JETP 20, 757 (1965)

\bibitem{BrownKibble_PR1964}
L.S. Brown, T.W.B. Kibble, Phys. Rev. 133(3A), A 705 (1964)

\bibitem{MarklundShukla_RMP2006}
M. Marklund, P.K. Shukla, Rev. Mod. Phys. 78, 591 (2006)

\bibitem{Ehlotzky_RPP2009}
F. Ehlotzky, K. Krajewska, J.Z. Kaminski, Rep. Prog. Phys. 72, 046401 (2009)

\bibitem{Ruffini_PR2010}
R. Ruffini, G. Vereshchagin, and S-S. Xue, Phys. Rep. 487, 1 (2010)

\bibitem{Piazza_RMP2012}
A. Di Piazza, C. Müller, K.Z. Hatsagortsyan, C.H. Keitel, Rev. Mod. Phys. 84(3), 1177 (2012)

\bibitem{ChengWu_PRL1969_22}
H. Cheng, T.T. Wu, Phys. Rev. Lett. 22, 666 (1969)

\bibitem{ChengWu_PRL1969_23}
 H. Cheng, T.T. Wu, Phys. Rev. Lett. 23, 1311 (1969)

\bibitem{ChengWu_PR1969_i}
 H. Cheng, T.T. Wu, Phys. Rev. 182, 1852 (1969)

\bibitem{ChengWu_PR1969_ii}
H. Cheng, T.T. Wu, Phys. Rev. 182, 1868 (1969)

\bibitem{ChengWu_PR1969_iii}
H. Cheng, T.T. Wu, Phys. Rev. 182, 1873 (1969)

\bibitem{ChengWu_PR1969_iv}
H. Cheng, T.T. Wu, Phys. Rev. 182, 1899 (1969)

\bibitem{ChengWu_PRD1970_p3414}
 H. Cheng, T.T. Wu, Phys. Rev. D 1, 13414 (1970)

\bibitem{ChengWu_PRD1970_p467}
H. Cheng, T.T. Wu, Phys. Rev. D 1, 467 (1970)

\bibitem{ChengWu_PRD1970_2}
 H. Cheng, T.T. Wu, Phys. Rev. D 2(9), 2103 (1970)

\bibitem{FriedEberly_PR1964}
Z. Fried, J.H. Eberly, Phys. Rev. 136, B 871 (1964)

\bibitem{Reiss_LP2005}
H.R. Reiss, Laser Physics 15(10), 1486 (2005)

\bibitem{Reiss_EPJD2009}
H.R. Reiss, Eur. Phys. J. D 55(2), 365 (2009)

\bibitem{Gordon_ZP1926}
 W. Gordon, Z. Phys. 40, 117 (1926)

\bibitem{Volkov_ZPhys1935}
D.M. Wolkow, Z. Phys. 94, 250 (1935)

\bibitem{Boca_JPA2011}
M. Boca, J. Phys. A: Math. Theor. 44, 445303 (2011)

\bibitem{Berestetskii_Elsevier1982}
V.B. Berestetskii, E.M. Lifshitz, V.B. Pitaevskii, Quantum Electrodynamics (Elsevier, New York,
1982)

\bibitem{SenGupta1952}
N.D. Sen Gupta, Bull. Math. Soc. (Calcutta) 44, 175 (1952)

\bibitem{Ritus_1985}
 V.I. Ritus, J. Sov. Laser Res. 6, 497 (1985)

\bibitem{Goldman_a}
 I. Goldman, J. Experimental Theoretical Phys. 19, 954 (1964)

\bibitem{Reiss_PQE1992}
 H.R. Reiss, Prog. Quant. Electr. 16, 1 (1974)


\bibitem{Korsch_JPA2006}
H.J. Korsch, A. Klumpp, D. Witthaut, J. Phys. A: Math. Gen. 39, 14947 (2006)

\bibitem{Dirac_RMP1949}
P.A.M. Dirac, Rev. Mod. Phys. 21, 392 (1949)

\end{thebibliography}

\end{document}